\documentclass[12pt]{iopart}
\usepackage{iopams}
\usepackage{epsfig}
\begin{document}

\title[Evolutionary dynamics of cooperation in neutral populations]{Evolutionary dynamics of cooperation in neutral populations}

\author{Attila Szolnoki$^1$ and Matja{\v z} Perc$^{2,3,4}$}
\address{$^1$Institute of Technical Physics and Materials Science, Centre for Energy Research, Hungarian Academy of Sciences, P.O. Box 49, H-1525 Budapest, Hungary\\
$^2$Faculty of Natural Sciences and Mathematics, University of Maribor, Koro{\v s}ka cesta 160, SI-2000 Maribor, Slovenia\\
$^3$CAMTP -- Center for Applied Mathematics and Theoretical Physics, University of Maribor, Krekova 2, SI-2000 Maribor, Slovenia\\
$^4$Complexity Science Hub, Josefst{\"a}dterstra{\ss}e 39, A-1080 Vienna, Austria}
\ead{szolnoki.attila@energia.mta.hu, matjaz.perc@uni-mb.si}

\begin{abstract}
Cooperation is a difficult proposition in the face of Darwinian selection. Those that defect have an evolutionary advantage over cooperators who should therefore die out. However, spatial structure enables cooperators to survive through the formation of homogeneous clusters, which is the hallmark of network reciprocity. Here we go beyond this traditional setup and study the spatiotemporal dynamics of cooperation in a population of populations. We use the prisoner's dilemma game as the mathematical model and show that considering several populations simultaneously give rise to fascinating spatiotemporal dynamics and pattern formation. Even the simplest assumption that strategies between different populations are payoff-neutral with one another results in the spontaneous emergence of cyclic dominance, where defectors of one population become prey of cooperators in the other population, and vice versa. Moreover, if social interactions within different populations are characterized by significantly different temptations to defect, we observe that defectors in the population with the largest temptation counterintuitively vanish the fastest, while cooperators that hang on eventually take over the whole available space. Our results reveal that considering the simultaneous presence of different populations significantly expands the complexity of evolutionary dynamics in structured populations, and it allow us to understand the stability of cooperation under adverse conditions that could never be bridged by network reciprocity alone.
\end{abstract}

\maketitle

\section{Introduction}
Methods of statistical physics, in particular Monte Carlo simulations and the theory of phase transitions \cite{stanley_71, marro_99, hinrichsen_ap00}, have been successfully applied to a rich plethora of challenging problems in the social sciences \cite{castellano_rmp09, helbing_jsp15, orsogna_plr15, pastor_rmp15}. The evolution of cooperation in social dilemmas --- situations where what is best for the society is at odds with what is best for an individual --- is a vibrant example of this development. Many reviews \cite{szabo_pr07, perc_bs10, perc_jrsi13, wang_z_epjb15} and research papers that reveal key mechanisms for socially preferable evolutionary outcomes have been published in recent years  \cite{zimmermann_pre05, santos_prl05, santos_pnas06, pacheco_prl06, gomez-gardenes_prl07, masuda_prsb07, fu_pre09, floria_pre09, lee_s_prl11, tanimoto_pre12, mobilia_pre12, szolnoki_epl16, fu2017leveraging, battiston2017determinants, szolnoki2017second}. Since cooperative behaviour is central to the survival of many animal species, and since it is also at the heart of the remarkable evolutionary success story of humans \cite{hrdy_11, nowak_11}, it is one of the great challenges of the 21st century that we succeed in understanding how best to sustain and promote cooperation \cite{axelrod_84}.

It has been shown that phase transitions leading to cooperation depend sensitively on the structure of the interaction network and the type of interactions \cite{albert_rmp02, boccaletti_pr06, holme_sr12, kivela_jcn14, boccaletti_pr14}, as well as on the number and type of competing strategies \cite{szabo_pr07,roca_plr09,szolnoki_jrsif14,perc_pr17}. An important impetus for the application of statistical physics to evolutionary social dilemmas and cooperation has been the seminal discovery of Nowak and May \cite{nowak_n92b}, who showed that spatial structure can promote the evolution of cooperation through the mechanism that is now widely referred to as network reciprocity \cite{nowak_s06, rand_pnas14}. A good decade latter Santos and Pacheco have shown just how important the structure of the interaction network can be \cite{santos_prl05}, which paved the way further towards a flourishing development of this field of research.

But while research concerning the evolutionary dynamics of cooperation in structured populations has come a long way, models where different populations do not interact directly but compete for space at the level of individuals have not been considered before. Motivated by this, we consider a system where two or more populations are distributed randomly on a common physical space. Between the members of a particular population the interactions are described by the prisoner's dilemma game. But there are no such interactions between players belonging to different populations, and hence players are unable to collect payoffs from neighbors belonging to a different population. The populations on the same physics space, for example on a square lattice, are thus neutral. Nevertheless, all players compete for space regardless to which population they belong, so that a player with a higher fitness is likely to invade a neighboring player with a lower fitness.

As we will show, such a conglomerate of otherwise neutral populations gives rise to fascinating spatiotemporal dynamics and pattern formation that is rooted in the spontaneous emergence of cyclic dominance. Within a very simple model, we observe the survival of cooperators under extremely adverse conditions where traditional network reciprocity would long fail, and we observe the dominance of the weakest due to the greediness of the strongest when considering different temptations to defect in different populations.

In what follows, we first present the studied prisoner's dilemma game and the details of the mathematical model. We then proceed with the presentation of the main results and a discussions of their wider implications.

\section{Prisoner's dilemma in neutral populations}
As the backbone of our mathematical model, we use a simplified version of the prisoner's dilemma game, where the key aspects of this social dilemma are preserved while its strength is determined by a single parameter \cite{nowak_n92b}. In particular, mutual cooperation yields the reward $R=1$, mutual defection leads to punishment $P=0$, while the mixed choice gives the cooperator the sucker's payoff $S=0$ and the defector the temptation $T>1$. We note that the selection of this widely used and representative parameterisation gives results that remain valid in a broad range of pairwise social dilemmas, including the snowdrift and the stag-hunt game.

All players occupy the nodes of a $L \times L$ square lattice with four neighbours each. To introduce the simultaneous presence of different populations, the $L^2$ players are assigned to $i=1,2, \ldots, n$ different populations uniformly at random. All $i$ populations contain an equal fraction of $C_i$ cooperators and $D_i$ defectors, who upon pairwise interactions receive payoffs in agreement with the above-described prisoner's dilemma game. Importantly, between different populations players are payoff-neutral with one another, which means that when $C_i$ meets $C_j$ or $D_j$, its payoff does not change, and vice versa. In the next subsection, we first consider the model where all populations have the same temptation to defect ($T_i=T$ for all $i$), and then we relax this condition to allow different temptations to defect in different populations.

We use the Monte Carlo simulation method to determine the spatiotemporal dynamics of the mathematical model, which comprises the following elementary steps. First, a randomly selected player $x$ acquires its payoff $\Pi_x$ by playing the game potentially with all its four neighbours. Next, player $x$ randomly chooses one neighbour $y$, who then also acquires its payoff $\Pi_y$ in the same way as previously player $x$. Finally, player $x$ imitates the strategy of player $y$ with the probability $w=\{1+\exp((\Pi_x-\Pi_y)/K)\}^{-1}$, where we use $K=0.1$ as the inverse of the temperature of selection to obtain results comparable with existing research \cite{szabo_pr07}. Naturally, when neighbouring players compete for space then the above describe microscopic dynamics involves not only the adoption of more successful strategy but also the imitation of the involved population tag.

In agreement with the random sequential simulation procedure, during a full Monte Carlo step (MCS) each player obtains a chance once on average to imitate a neighbor. The average fractions of all microscopic states on the square lattice are determined in the stationary state after a sufficiently long relaxation time. Depending on the proximity to phase transition points and the typical size of emerging spatial patterns, the linear system size was varied from $L=400$ to $6600$, and the relaxation time was varied from $10^4$ to $10^6$ MCS to ensure that the statistical error is comparable with the size of the symbols in the figures.

\section{Results}

\begin{figure}
\centering
\includegraphics[width=0.6\linewidth]{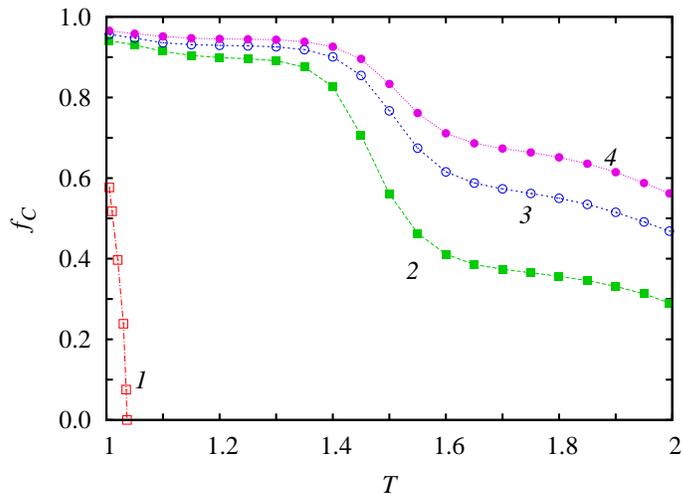}
\caption{The stationary $f_C$ fraction of cooperators in the whole system in dependence on the temptation to defect $T$, as obtained for different numbers of populations $n$ that form the global ssystem (indicated by the number along each curve). For reference the result of the classic one-population ($n=1$) spatial prisoner's dilemma game is shown as well. These results indicate that the introduction of additional populations whose members are payoff-neutral between one another significantly improves the survival chances of cooperators.}
\label{sym}
\end{figure}

Naively, one might assume that introducing several populations simultaneously which bear the same serious conflict of competing strategies might not bring about any changes in the evolutionary outcome. As is well known, the Nash equilibrium of the prisoner's dilemma game is mutual defection \cite{nash_am51}, and since this applies to all populations, the overall outcome should be mutual defection too. This reasoning is actually completely correct in well-mixed populations, where the consideration of different, otherwise neutral populations really does not change the result: cooperators die out in all populations as soon as $T>1$. But as we will show next, this naive expectation is completely wrong in structured populations, where excitingly different evolutionary outcomes can be observe due to the simultaneous presence of different populations.

\begin{figure}
\centering
\includegraphics[width=0.7\linewidth]{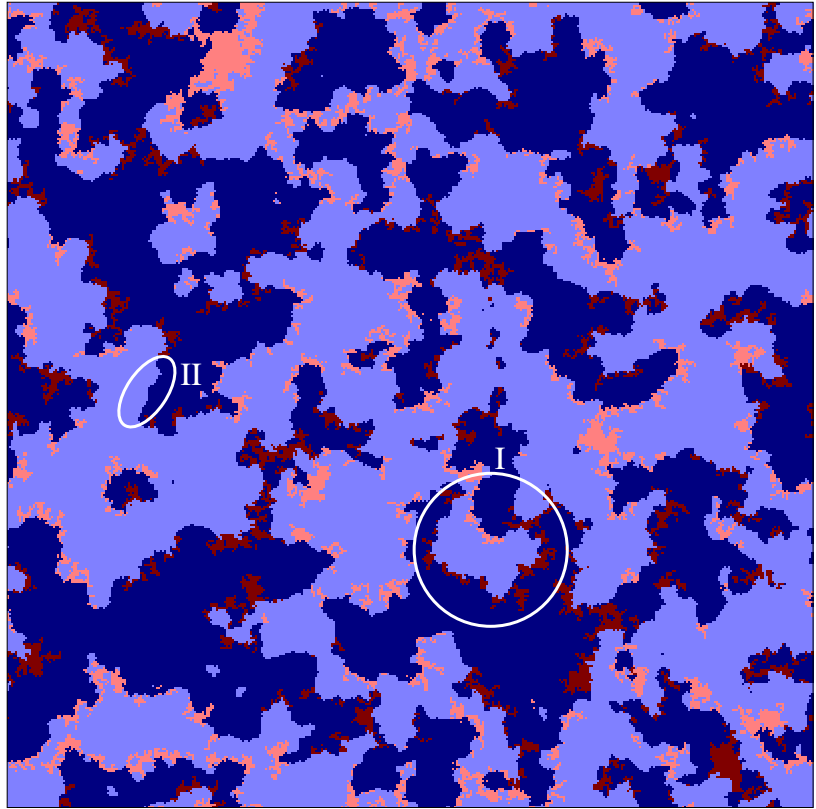}
\caption{Snapshot of the distribution of strategies in the stationary state in a system consisting of two populations. In both population the temptation to defect is $T=1.2$. Defectors belonging to the first population are depicted dark red, while cooperators of the first population are depicted dark blue. Similarly, cooperators and defectors of the second population are depicted light blue and light red, respectively. The key mechanism that is responsible for the emerging spatial pattern is highlighted by a white circle marked `I'. Together with the animation provided in \cite{2-rand}, it can be observed that dark red defectors invade dark blue cooperators, but light blue cooperators invade dark red defectors. Likewise, light red defectors invade light blue cooperators, but dark blue cooperators invade light red defectors. This spontaneous emergence of cyclic dominance in the form $D_1 \to C_1 \to D_2 \to C_2 \to D_1$ is responsible for the sustenance of cooperation even at very high temptation values that can be observed in Fig.~\ref{sym}. The white ellipse marked `II' highlights the smooth interface between both cooperator strategies in the absence of defectors, which is surprising given that the two strategies are payoff-neutral and thus should be subject to voter-model-like coarsening. For clarity a $L=400$ linear system size was used.}
\label{snapshot}
\end{figure}

As far as cooperation promotion is concerned, and before elucidating the responsible microscopic mechanism for such favourable evolutionary outcomes, we show in Fig.~\ref{sym} how the fraction of cooperators changes in dependence on the temptation to defect $T$ for different numbers of populations $n$ that form the global system. For comparison, we also show the baseline $n=1$ case, which corresponds to the traditional version of the weak prisoner's dilemma game on the square lattice, and where cooperators benefit from network reciprocity to survive up to $T \leq 1.037$ \cite{szabo_pre05}. It can be observed that, as we increase $n$, the fraction of cooperators increases dramatically. In fact, the higher the value of $n$, the higher the stationary fraction of cooperators in the whole system.

The spatiotemporal dynamics behind this promotion of cooperation in a complex system consisting of two populations can be seen in the animation provided in \cite{2-rand}, while a representative snapshot of the stationary state is shown in Fig.~\ref{snapshot}. In both cases cooperators are depicted blue while defectors are depicted red, and different shades of these two colours denote adherence to the two different populations. In Fig.~\ref{snapshot}, we have circled two crucial details that explain how the patterns evolve over time. The white circle marked `I' highlights that dark red defectors can easily invade dark blue cooperators. However, the invaded space is quickly lost to light blue cooperators belonging to the other population. The latter, on the other hand, are successfully invaded by light red defectors from their own population, who are in turn again invaded by dark blue cooperators. In this way the loop is closed, revealing the spontaneous emergence of cyclic dominance in the form $D_1 \to C_1 \to D_2 \to C_2 \to D_1$, which determines the stationary distribution of strategies in our system. As is well-known, the cyclic dominance is crucial for the maintenance of biodiversity \cite{szolnoki_jrsif14}, which in our case translates to the survival of all four competing strategies, and thus to the sustenance of cooperation even at very high temptation values.

\begin{figure}
\centering
\includegraphics[width=.7\linewidth]{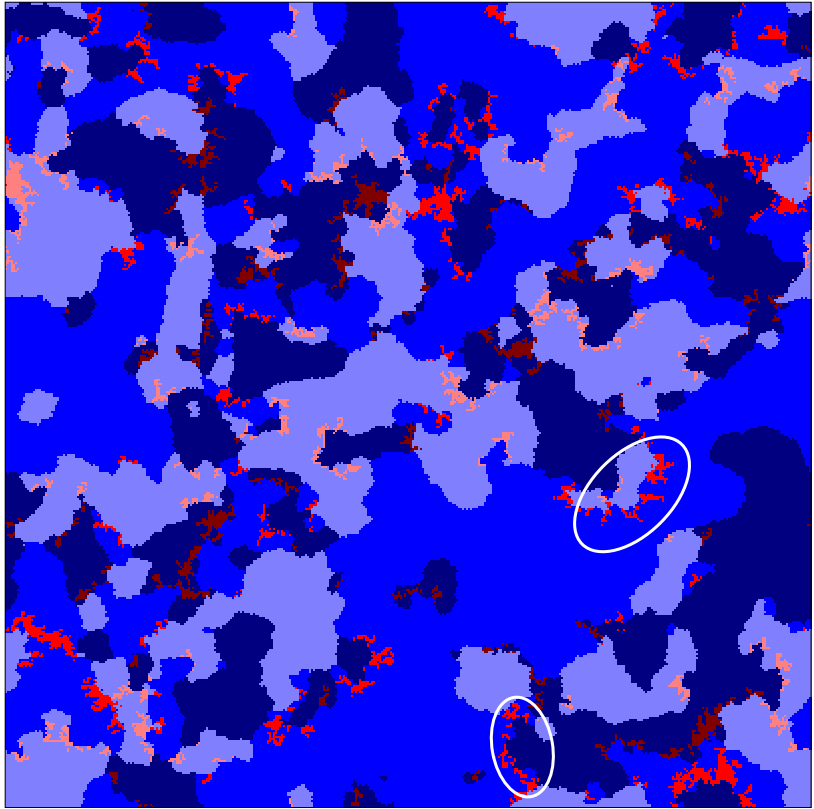}
\caption{Snapshot of the distribution of strategies in the stationary state in a system consisting of three populations. In all three population the temptation to defect is $T=1.2$. As in Fig.~\ref{snapshot}, different shades of blue and red depict cooperators and defector belonging to different populations. White ellipses highlight that plain red defectors are successfully invaded by both light blue and dark blue cooperators that belong to the other two populations.}
\label{three_s}
\end{figure}

This cyclic dominance can be observed directly if we launch the evolution from a prepared initial state, such that homogeneous domains of the competing strategies are separated by straight interfaces, as in the animation provided in \cite{2-patch} (in this animation a higher $T=1.5$ temptation to defect was used to yield clearer propagating fronts). It can be observed that conceptually similar propagating fronts emerge as were observed before in rock-paper-scissors-like systems \cite{szolnoki_jrsif14}.

Turning back to Fig.~\ref{snapshot}, the white ellipse marked `II' highlights another important aspect of the spatiotemporal dynamics, namely the smooth interface separating the two cooperative strategies in the absence of defectors. This may be surprising at first because these strategies are payoff-neutral, and thus a voter-model-like coarsening with highly fluctuating interfaces would be expected \cite{dornic_prl01}. Indeed, while a $C_1$ cooperator does not benefit from the vicinity of a $C_2$ cooperator, other $C_1$ cooperators close by of course increase each other's payoffs (and vice versa for $C_2$ cooperators). As a consequence of this the payoffs of $C_1$ and $C_2$ cooperators along the interface differ, so that one will likely invade the other. This process always aims to straighten the interfaces. If an interface cannot be straightened, for example around a small island, the latter will shrink due to an effective surface tension.

Lastly in terms of the results presented in Fig.~\ref{sym}, it remains to explain why the larger the number of populations forming the global system the higher the level of cooperation in the stationary state, and this regardless of the temptation to defect. To that effect we provide in \cite{3-rand} an animation showing the spatiotemporal dynamics when $n=3$, and in Fig.~\ref{three_s} a representative snapshot of the distribution of strategies on the square lattice in the stationary state. These results reveal that the increasing positive effect is due to the fact that the addition of one new population $i$ always yields one additional prey to the cooperators in other populations. At the same time, no new predators to them are introduced, i.e., $D_i$ defectors who act as the prey to cooperators in the other populations are predators only to $C_i$ cooperators, but the latter find their prey in defectors from other populations too. The snapshot in Fig.~\ref{three_s} features two white ellipses, where it is highlighted that the plain red $D_3$ defectors are dominated by both $C_1$ (dark blue) and $C_2$ (light blue) cooperators (see also the animation in \cite{3-rand}).

Thus far, we have only considered cases where the temptation to defect was the same in all populations. By relaxing this restriction, the number of free parameters increases significantly, yet it is still possible to determine general properties of the spatiotemporal dynamics that governs the evolutionary outcomes in a presented system.

\begin{figure}
\centering
\includegraphics[width=0.6\linewidth]{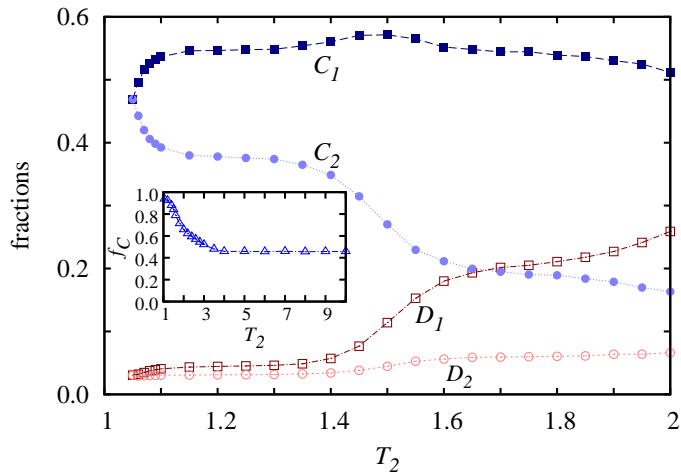}
\caption{The stationary fractions of the four competing strategies in dependence on the temptation to defect $T_2$, as obtained for $n=2$. The temptation to defect for the first population is $T_1=1.05$. The inset shows the overall fraction of cooperators in the system in the large $T_2$ limit.}
\label{asym2}
\end{figure}

We begin by presenting results for the generalized two-population setup where $T_1 \neq T_2$. As we have shown above, the emergence of cyclic dynamics between the four competing microscopic states in general dictates a stable coexistence. By increasing the temptation to defect in one population practically increases the rate in the corresponding $D \to C$ invasion. The consequences of this fact, based on the fundamental principles of cyclic dominance \cite{szolnoki_jrsif14}, actually completely explain the evolutionary outcomes in Fig.~\ref{asym2}. The first potentially surprising observation is that increasing the temptation to defect $T_2$ between $D_2$ defectors and $C_2$ cooperators will not only lower the stationary fraction of $C_2$ and increase the stationary fraction of $D_2$, but also elevate the fraction of $C_1$ cooperators. This is because $D_2$ defectors are prey to $C_1$ cooperators, and it is well-known that a species entailed in cyclic dominance is promoted not by weakening its predator, but rather by making its prey stronger. This paradox is a frequently observed trademark of systems that are governed by cyclic dominance \cite{frean_prsb01}. However, despite the described boost to the growth of $C_1$ cooperators, the overall fraction of all cooperators in the whole system decreases slightly as we increase $T_2$ towards very large values, as illustrated in the inset of Fig.~\ref{asym2}.

For a better demonstration of the acceleration of the $D_2 \to C_2$ invasion and the resulting boost to $C_1$ cooperators (dark blue), we provide an animation in \cite{2-large}, where an extreme high $T_2=100$ was used at $L=400$ linear system size. As the animation shows, although $C_2$ cooperators (light blue) are invaded very efficiently by $D_2$ defectors (light red), the abundance of $D_1$ defectors (dark red) always offers an evolutionary escape hatch out of extinction of $C_2$ cooperators. In agreement with the above described cyclic dominance, $D_2$ defectors are fast invaded by $C_1$ cooperators. Interestingly, $D_1$ defectors would also beat $C_1$ cooperators because $T_1=1.05$ is above the $T=1.037$ cooperation survival threshold of a single population, yet the $D_2 \to C_2$ propagating front always comes to the rescue, bringing with it $D_2$ defectors as prey.

\begin{figure}
\centering
\includegraphics[width=0.7\linewidth]{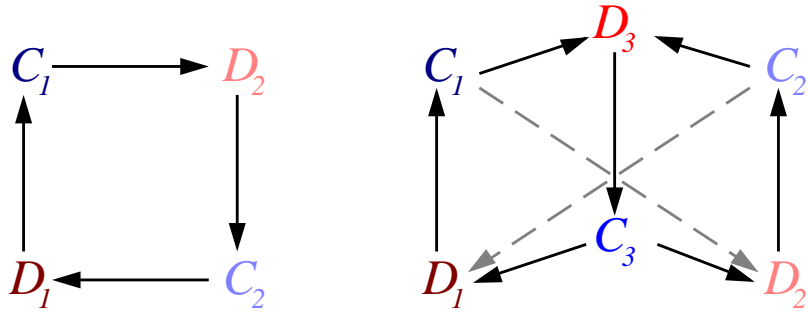}
\caption{The effective food-web of all competing strategies in a two- (left) and three-population (right panel) system. We emphasize that the depicted relations between strategies exist only in a spatial system, where cooperators can invade defectors from other populations. If we consider solely pairwise interactions, the relation between $C_1$ cooperators and $D_2$ defectors (or $C_1$ and $C_2$) is of course payoff-neutral, as defined in the mathematical model.}
\label{web}
\end{figure}

In comparison to the results obtained when the temptation to defect is the same in all populations (see Fig.~\ref{sym}), it may come as a surprise that cooperators die out if $T > 2.85$, and this despite the fact that qualitatively the same cyclic dominance emerges there. The explanation of this difference illustrated in Fig.~\ref{asym2} is that in the symmetrical case the $D_1 \to C_2$ and $D_2 \to C_2$ invasion rates change simultaneously as we vary $T$. However, it is precisely this simultaneous change of invasion rates that may jeopardize the stable coexistence in models of cyclic dominance. As shown previously for a symmetric $4-$strategy Lotka-Volterra system, the coexistence disappears if the difference between the invasion rates exceeds a threshold value \cite{szabo_pre08}. For an illustration, the effective food-web of the four competing strategies in a two-population model is shown in left panel of Fig.~\ref{web}.

Naturally, if we allow different temptation values in different populations the behaviour becomes even more complex, as we show next using still a relatively simple three-population system as an example. The effective food-web is shown in the right panel of Fig.~\ref{web}. If we just vary $T_3$, while the temptation to defect in the other two populations remains fixed at $T_1=T_2=1.05$, the $D_3 \to C_3$ invasion rate will influence invasions in several other cycles in the effective food-web. Examples include the $D_3 \to C_3 \to D_1 \to C_1 \to D_3$ cycle, or the $D_3 \to C_3 \to D_2 \to C_2 \to D_3$ cycle, or the $D_3 \to C_3 \to D_1 \to C_1 \to D_2 \to C_2 \to D_3$ cycle, all of which contain the elementary $D_3 \to C_3$ invasion that is directly affected by $T_3$. This is why it is almost impossible to predict the response of a system comprised of several neutral populations, even if only a single temptation to defect is varied.

\begin{figure}
\centering
\includegraphics[width=0.6\linewidth]{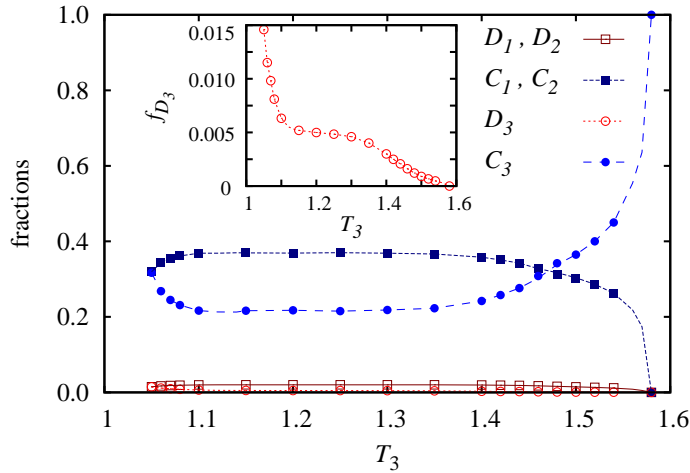}
\caption{The stationary fractions of the six competing strategies in dependence on the temptation to defect $T_3$, as obtained for $n=3$. The temptation to defect for the first and second population is $T_1=T_2=1.05$. The inset shows the fraction of $D_3$ defectors in dependence on $T_3$. Counterintuitively, although larger $T_3$ values directly support $D_3 \to C_3$ invasions, the fraction of $D_3$ defectors decreases steadily as the value of $T_3$ increases.}
\label{asym3}
\end{figure}

For the above $n=3$ case, the results showing how different $T_3$ values affect the evolutionary outcome are presented in Fig.~\ref{asym3}. It can be observed that upon increasing the value of $T_3$, the stationary fraction of $C_1$ and $C_2$ cooperators is not affected, even though they are the predators of $D_3$ who should in principle be promoted by large $T_3$ values. On the other hand, the overall fraction of all defectors in the system remains very low. But the most exotic reaction is that of the fraction of $C_3$ cooperators, which is of course the direct prey of $D_3$ defectors. While initially their fraction in the stationary state decreases to a shallow minimum across the intermediate range of $T_3$ values, it ultimately increases to complete dominance above a threshold value. In other words, while defectors survive when all $T$ values in the system are equal to $1.05$, they die out if we increase one of them sufficiently, as it happens in Fig.~\ref{asym3} when the $T_3$ value is sufficiently large. Due to the symmetry of the model the same results are of course obtained if either the value of $T_1$ or $T_2$ would be enlarged instead of the value of $T_3$.

\begin{figure}
\centering
\includegraphics[width=.7\linewidth]{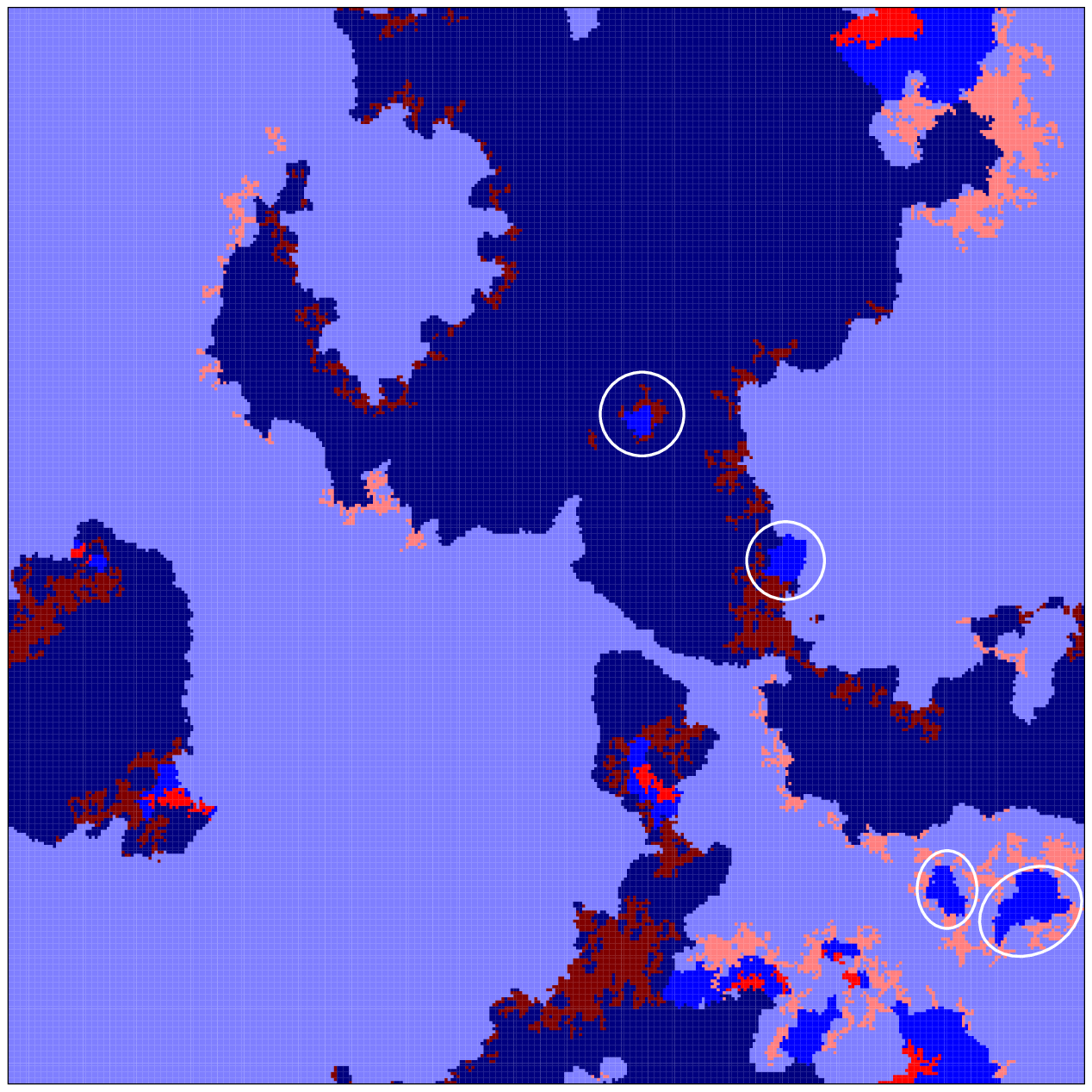}
\caption{Snapshot of the distribution of strategies during an early stage of evolution in a system consisting of three populations. Values of the temptation to defect are $T_3=1.8$ and $T_1=T_2=1.05$. As in Figs.~\ref{snapshot} and \ref{three_s}, different shades of blue and red depict cooperators and defector belonging to different populations. White ellipses highlight the weakest $C_3$ cooperators (plane blue), who manage to survive despite the large $T_3$ value giving a huge evolutionary advantage to their direct predators $D_3$ (plain red). What is more, due to their greediness, $D_3$ defectors are actually the first to die out, thus paving the way for $C_3$ cooperators to rise to complete dominance by using $D_1$ and $D_2$ defectors (light and dark red) as a Trojan horse to invade the territory of $C_1$ and $C_2$ cooperators (light and dark blue). This is an example where the weakest ultimately dominate because of the greediness of the strongest. For clarity a $L=360$ linear system size was used.}
\label{three_as}
\end{figure}

To better understand and illustrate the seemingly paradoxical effect the increasing $T_3$ value has on the evolutionary outcome, we provide an animation from a prepared initial state in \cite{3-prep}. Here the square lattice is horizontally divided into two parts, where in the top half $C_1$ cooperators (dark blue) are framed by $D_1$ defectors (dark red), while in the bottom half $C_3$ cooperators (plane blue) are framed by $D_3$ defectors (plain red). Moreover, both $D-C-D$ patches are surrounded by $C_2$ cooperators (light blue). The applied  temptation to defect values are $T_1=T_2=1.05$ and $T_3=1.8$. Importantly, invasions through the horizontal border are not permitted because we want to compare the independent evolution of both sub-systems. Since $D_2$ defectors are not present, $C_2$ cooperators have no natural predator. As a consequence, the whole system will evolve into a pure $C_2$ (light blue) phase. However, the really interesting aspect of this animation is how the mentioned sub-systems reach this state. In the top half, $D_1$ defectors are less aggressive, and therefore their invasions are less salient. This has two important consequences. In the first place, their payoffs are not that high for the other strategies to imitate them, and so the $C_1-D_1$ border is fluctuating rather strongly. Secondly, $D_1$ defectors do not form a homogeneous front along this border. The latter would be essential for a fast invasion of $C_2$ cooperators (light blue), who are their predators. In other words, the effective invasion of $C_2$ cooperators can only happen via the invasion of $D_1$ defectors. The latter conditions is completely fulfilled in the bottom half where $D_3$ defectors are more aggressive. Here defectors form not just a more compact invasion front, but they also form a thick, uniform stripe, which is an easy target for $C_2$ cooperators. Consequently, the more aggressive defectors will die out much faster than their less potent $D_1$ counterparts in the top half of the square lattice.

This process just described is actually very common when the value of the temptation to defect in one population is significantly larger than the corresponding values in other populations. Of course, the extinction of the most aggressive defector frequently involves also the extinction of its cooperator prey. Sometimes, however, if the system size is large enough, it may happen that the prey of the more aggressive defectors manages to separate itself in an isolated part of the lattice and hang on until his predators die out. Such a situation is illustrated in Fig.~\ref{three_as}, where the white ellipses and circles mark plain blue cooperator spots who got rid of their natural predators (plain red). In the absence of the latter, the arguably weakest cooperators become the strongest, and they eventually rise to complete dominance by invading defectors from the other two populations who themselves continuously invade their cooperators. The whole evolutionary process can be seen in the animation in \cite{3-patch}, where we have used prepared initial patches of the six competing strategies to make the spatiotemporal dynamics that leads to the described pattern formation better visible. Additionally, for a faster evolution, we have used a smaller $L=180$ linear system size. In effect, the plain blue cooperators use the defectors from the other two populations as a Trojan horse to invade the whole available space. And despite of starting as the weakest, they turn out to be the dominant due to the greediness of their direct predators.

\section{Discussion}
We have studied the spatiotemporal dynamics of cooperation in a system where several neutral populations are simultaneously present. The evolutionary prisoner's dilemma game has been used as the backbone of our mathematical model, where we have assumed that strategies between the populations are payoff-neutral but competing freely with one another as determined by the interaction graph topology. Within a particular population the classical definition of the prisoner's dilemma game between cooperators and defectors has been applied. We have observed fascinating spatiotemporal dynamics and pattern formation that is unattainable in a single population setup. From the spontaneous emergence of cyclic dominance to the survival of the weakest due to the greediness of the strongest, our results have revealed that the simultaneous presence of neutral populations significantly expands the complexity of evolutionary dynamics in structured populations. From the practical point of view, cooperation in the proposed setup is strongly promoted and remains viable even under extremely adverse conditions that could never be bridged by network reciprocity alone. The consideration of simultaneously present neutral populations
thus allows us to understand the extreme persistence and stability of cooperation without invoking strategic complexity, and indeed in the simplest possible terms as far as population structure and overall complexity of the mathematical model is concerned.

The central observation behind the promotion of cooperation is that, if we put two payoff-neutral populations together, then only cooperators can benefit from it in the long run. While the advantage of mutual cooperation is readily recognizable already in a single population, and it is in fact the main driving force behind traditional network reciprocity, the extend of it remains limited because cooperators at the frontier with defectors always remain vulnerable to invasion. This danger is here elegantly avoided when a cooperative cluster meets with the defectors of the other population. In the latter case the positive consequence of network reciprocity is augmented and cooperators can easily invade the territory of the foreign defectors. Importantly, this evolutionary success of cooperators in one population works vice versa for cooperators in the other population(s) too. Due to this symmetry, it is easy to understand that, as we have shown, the larger the number of populations forming the system, the more effective the promotion of cooperation.

We have also shown that the already mentioned positive impact can be enhanced further if we allow different temptations to defect in different populations. Counterintuitively, in a system where the population specific temptation to defect values are diverse enough, defectors die out first whose temptation value is the largest. And this turns out to be detrimental for defectors in other populations too. An extreme aggressive invasion namely leads to the fast depletion of the prey -- in this case the cooperators from the corresponding population -- which in turn leads to the extinction of the predators. However, the reverse situation is not valid: if the most vulnerable cooperators somehow manage to survive, they eventually rise to complete dominance, using defectors from other populations as Trojan horses to invade cooperators from other populations. This gives rise to the dominance of the weakest due to the greediness of the strongest, and it also reminds us that dynamical processes in different populations should not be too diverse because this jeopardizes the stability of the whole system.

\ack
This research was supported by the Hungarian National Research Fund (Grant K-120785) and the Slovenian Research Agency (Grants J1-7009 and P5-0027).

\section*{References}
\providecommand{\newblock}{}

\end{document}